\documentclass[twocolumn,nofootinbib,superscriptaddress,aps]{revtex4-1}

\usepackage[T1]{fontenc} \usepackage{graphicx} \usepackage{amsmath}
\usepackage{amssymb} \usepackage{hyperref}

\newcommand{\dd}{\text{d}}
\begin{document}

\title{Thermal crumpling of perforated two-dimensional sheets}

\author{David~Yllanes}\email{dyllanes@syr.edu}\affiliation{Department of Physics and Soft Matter
Program, Syracuse University, Syracuse, NY, 13244} \affiliation{Kavli Institute
for Theoretical Physics, University of California, Santa Barbara, CA 93106,
USA} \affiliation{Instituto de Biocomputaci\'on y F\'{\i}sica de Sistemas
Complejos (BIFI), 50009 Zaragoza, Spain}
\author{Sourav S.~Bhabesh}\affiliation{Department of Physics and Soft Matter
Program, Syracuse University, Syracuse, NY, 13244} \affiliation{Kavli Institute
for Theoretical Physics, University of California, Santa Barbara, CA 93106,
USA} \author{David R.~Nelson}\affiliation{Department of Physics and School of
Engineering and Applied Sciences, Harvard University, Cambridge, MA 02138, USA}
\author{Mark~J.~Bowick}\affiliation{Department of Physics and Soft Matter
Program, Syracuse University, Syracuse, NY, 13244} \affiliation{Kavli Institute
for Theoretical Physics, University of California, Santa Barbara, CA 93106,
USA} \date{\today}

\begin{abstract} Thermalized elastic membranes without distant self-avoidance
are believed to undergo a crumpling transition when the microscopic bending
stiffness is comparable to $kT$, the scale of thermal fluctuations.
Most potential physical realizations of such
membranes have a  bending stiffness well in excess of experimentally achievable
temperatures  and are therefore unlikely ever to access the crumpling regime.
We propose a mechanism to tune the onset of the crumpling transition by
altering the geometry and topology of the sheet itself. We carry out
extensive molecular dynamics simulations of perforated sheets with a dense
periodic array of holes and observe that the critical temperature is
controlled by  the total fraction of removed area, independent of the precise
arrangement and size of the individual holes. The critical exponents for the
perforated membrane are compatible with those of the standard crumpling
transition.  \end{abstract}

\maketitle
Two-dimensional materials such as graphene~\cite{Katsnelson2012} or
MoS$_2$~\cite{Wang2012} currently enable the experimental
study~\cite{Nicholl2015} of the mechanical properties of thermalized elastic
sheets and a testing ground for many longstanding theoretical and numerical
predictions~\cite{Nelson1987,Aronovitz1988,Guitter1989,
LeDoussal1992,Zhang1993,Bowick1996,Bowick2001,Fasolino2007,Bowick2009,Gazit2009,
Los2009,Zakharchenko2010,Braghin2010,Hasselmann2011, Troster2013,Troster2015,
Fasolino2016, Kosmrlj2016}. Particularly striking is the possibility of
engineering elastic parameters such as the bending rigidity and Young's modulus
over broad ranges simply by varying the overall size or temperature of
atomically thin cantilevers and springs  (see \cite{Nelson2002,Nelson2004} for
general references on elastic membranes). 

Recent work by Blees et al.~\cite{Blees2015}, in addition to demonstrating a
4000-fold enhancement of the bending rigidity relative to its $T=0$ value, has
shown the potential of graphene as the raw ingredient of microscopic mechanical
metamaterials.  Employing the principles of kirigami (the art of
cutting paper), one can construct robust microstructures, thus providing an
alternative route to tune mechanical properties, leaving graphene's remarkable
electrical properties essentially intact.  These results serve as inspiration
for further theoretical work on the interplay between geometry and
mechanics~\cite{Russell2015}.

A common working model of elastic membranes is the crystalline or polymerized
membrane~\cite{Bowick2001}, which serves as a useful tool to describe systems
ranging from graphene to the spectrin cytoskeleton of red blood
cells~\cite{Schmidt1993}, polymersomes~\cite{Shum2008} and assemblies of spider
silk proteins~\cite{Hermanson2007}. In this context, a key theoretical
prediction is the existence of a crumpling transition. For low
temperatures, the thermal sheet is in a flat phase roughened by flexural
phonons, with long-range order in the orientations of surface normals,
analogous to the ferromagnetic phase of a spin system. At sufficiently high
temperatures, however, thermal fluctuations can disorder the membrane and drive
it to a crumpled phase, with only short-range order in the normals. This
prediction has been confirmed analytically and numerically, at least for
phantom membranes, without distant
self-avoidance~\cite{Bowick2001,Nelson2004, Kantor1987, Bowick1996}.
Unfortunately, in many materials, the crumpling transition is very far from the
experimentally accessible regime: for instance, graphene has a bending rigidity
of $\kappa_0 \approx 48kT_\text{R}$, where $T_\text{R}$ is room
temperature~\cite{Fasolino2007}.  For the intact lattice, this corresponds to a
crumpling temperature of order $10^4$-$10^5$K, well beyond the melting point
for graphene.  Clearly, we need a mechanism to lower the crumpling transition
temperature. 

 \begin{figure}[b] \begin{minipage}{.48\linewidth}
\includegraphics[width=\linewidth]{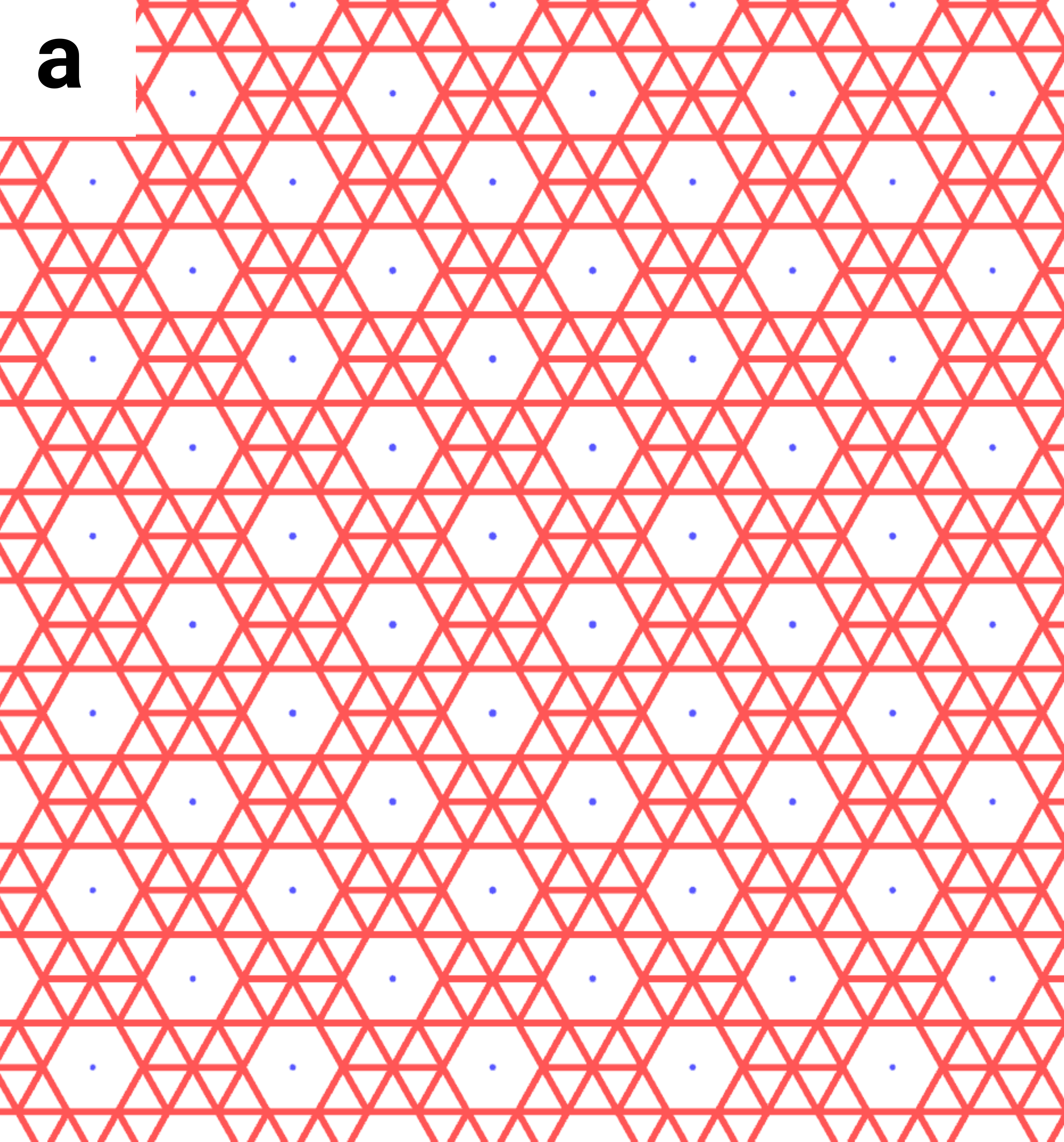} \end{minipage}
\begin{minipage}{.48\linewidth}
\includegraphics[width=\linewidth]{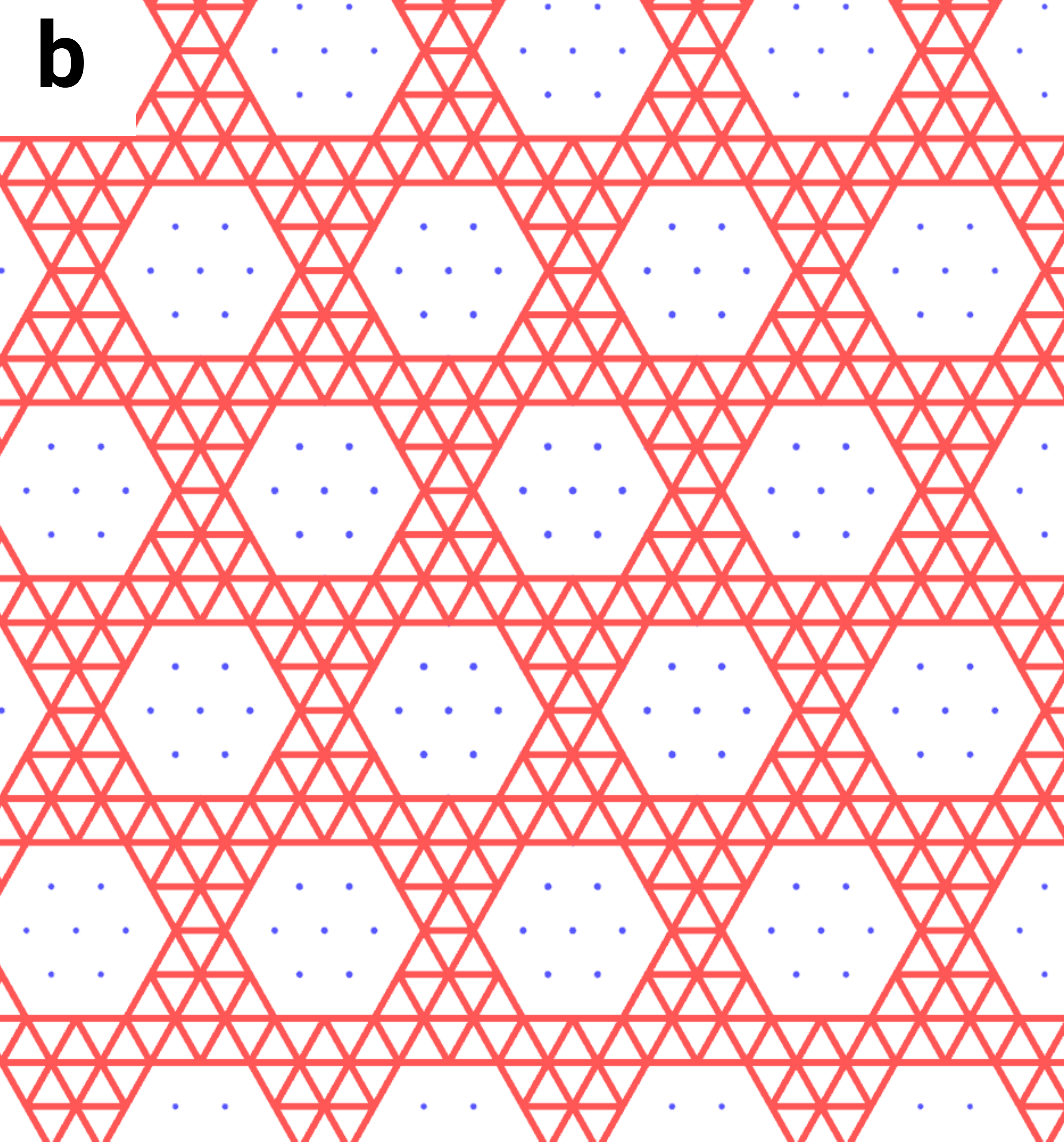} \end{minipage} 
\caption{Two different arrays of perforations. For clarity, the images show
only a small section of the full membrane. We
begin by considering a full triangulated sheet and then remove all the nodes in
a radius $R$ around its center. This central hole is then repeated periodically
throughout the membrane. In the figure the removed nodes are represented by
blue dots (\textbf{a:} $R=1$, \textbf{b:} $R=2$).  In the rest of the paper
we will consider patterns of perforations with $R=1,2$ and varying spacing
between holes (see Supplementary Fig.~1 for a full description of all
perforation patterns). \label{fig:pattern}} \end{figure}

We show here that the crumpling temperature can be significantly lowered by
altering the geometry and topology of the membrane. In particular, we perform
extensive molecular dynamics simulations of crystalline membranes with dense
periodic arrays of holes and determine the dependence of the onset of crumpling
on the degree of perforation. This dependence is very strong, but can be shown
to be a function of a simple control parameter, namely the total fraction of
removed area, and independent of the detailed arrangement and size of the
individual holes.  
\begin{figure}[t] \includegraphics[width=\linewidth]{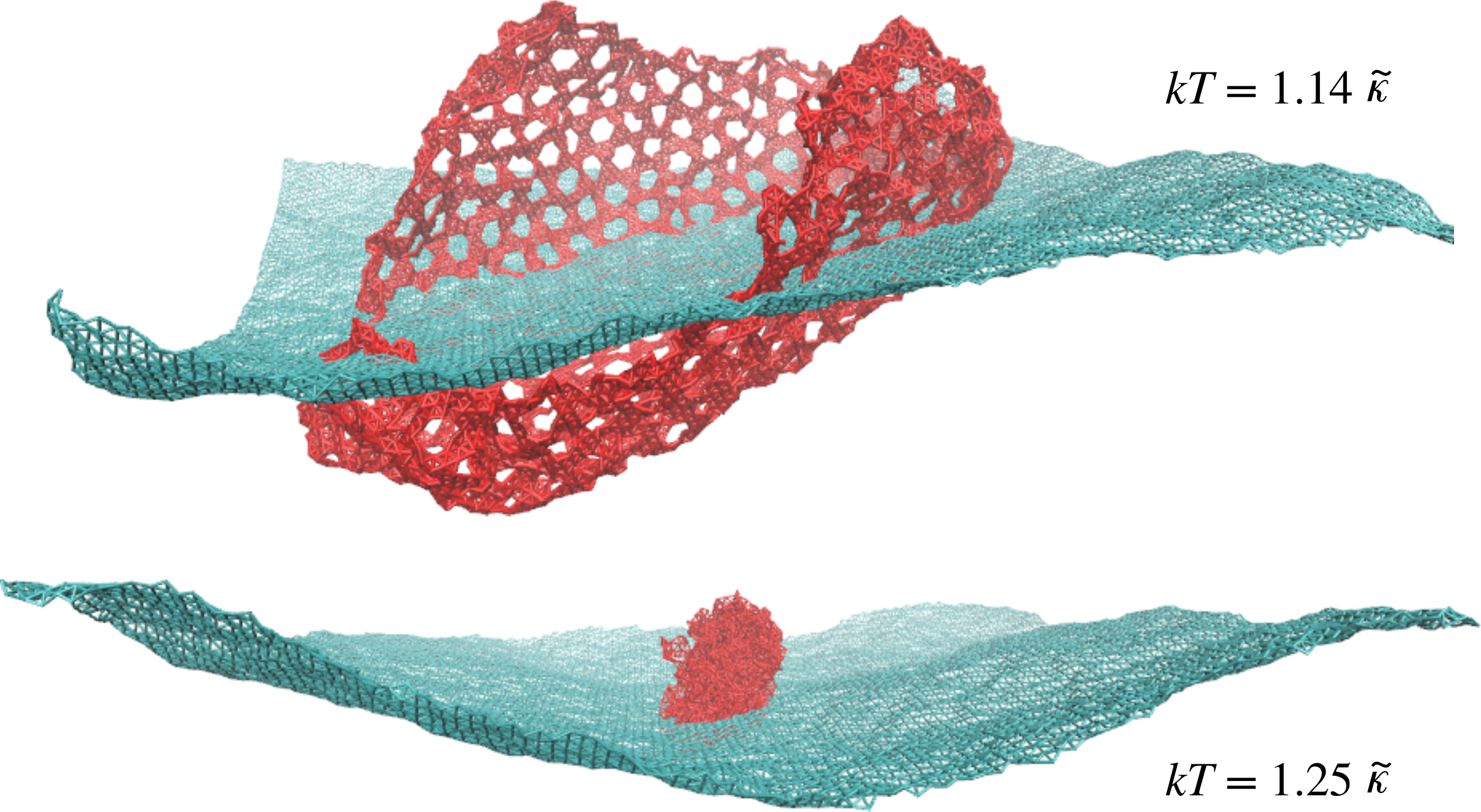}
\caption{Snapshots of thermalized configurations.  
We superimpose the configurations of a pristine
sheet (blue) and a perforated sheet with holes of size $R=2$, in the pattern of
Fig.~\ref{fig:pattern}-b (red) for two values of the temperature (in units
of the bending rigidity $\tilde \kappa$).  In both cases the full sheet is well
into the flat phase, the thermal fluctuations causing just some wrinkling and
oscillation. This $10\%$ increase in temperature, however, triggers a crumpling
of the perforated sheet. Both systems have size $L=100a$.  See Supplementary
Movies 1~and~2 for animations of these simulations.  \label{fig:crumpling}}
\end{figure}

\subsection*{Results}
\subsubsection*{Our model}  We consider square sheets of size $L\times L$, which for
the purposes of computer simulation we discretize with a tiling of equilateral
triangles of side $a=1$. We use a standard coarse-grained
model~\cite{Seung1988} to compute the elastic energy in the sheet, which is
composed of a stretching and a bending term \begin{equation}\label{eq:H}
\mathcal H = \mathcal H_\text{stretch} +  \mathcal H_\text{bend}.
\end{equation} Stretching is modeled by considering each triangle side as a
spring of elastic constant $\epsilon$ and rest length $a$: \begin{align}
\mathcal H_\text{stretch} = \frac12 \epsilon \sum_{\langle i,j\rangle}
(r_{ij}-a)^2, \label{eq:stretch} \end{align} where the sum is over all pairs of
vertices joined by a triangle edge. The bending energy is represented by a
standard dihedral interaction between normals, \begin{equation}\label{eq:bend}
\mathcal H_\text{bend} = \tilde \kappa \sum_{\langle \alpha,\beta\rangle}
(1-\hat{\mathbf n}_\alpha\cdot \hat{\mathbf n}_\beta).  \end{equation}
Here the sum is over all the pairs of triangles that share a side and
$\hat{\mathbf n}_\alpha$ is the unit normal to triangle
$\alpha$. Note that placing a carbon atom at the center of each
triangle provides an approximate atomic model for the elastic modes of graphene
on a dual lattice, as long as we choose the bending rigidity and Young's
modulus correctly.

The elastic parameters $\epsilon$  and $\tilde \kappa$ are directly related to
the continuum Young's modulus ($Y_0 = 2\epsilon/\sqrt3$) and bare bending
rigidity ($\kappa_0 = \sqrt3 \tilde \kappa/2$). Normally, when performing a
numerical study (see, e.g., \cite{Bowick1996,Cuerno2016}) one chooses natural
units where $\epsilon=1$ and $\tilde\kappa$ is varied. To better approximate
the behavior of materials such as graphene or MoS$_2$, however, we will instead
fix the ratio $\epsilon/\tilde\kappa$ and vary temperature by changing the
$\tilde\kappa/kT$ ratio.  For graphene at room temperature
$\tilde\kappa/kT\approx 48$~\cite{Fasolino2007}.
Following~\cite{Russell2015,Bowick2017}, we will use
$\epsilon/\tilde\kappa=1440/a^2$. This corresponds to a Young's modulus about
an order of magnitude lower than for real graphene~\cite{Lee2008,Zhao2009}, in
order to facilitate equilibration in our computer simulations. We note,
however, that this choice should have only a minor effect on the onset of
crumpling, since the degree of order in the normals only depends on
$\epsilon/\tilde\kappa$ logarithmically~\cite{Kosmrlj2016}.  For a sheet of
length $L/a=\mathcal O(10^2)$, corresponding to a patch of freely suspended
graphene roughly $30$nm on a side, these parameters result in a F\"oppl von
K\'arm\'an number of $\text{vK} = Y_0L^2/\kappa_0 \sim \mathcal O(10^7)$,
similar to that of a standard A4 sheet of paper.

\subsubsection*{Dense arrays of holes}

We are interested in exploring the effect of a perforated geometry on the
rigidity of elastic membranes. To this end we shall compare the physics of the
``full'' or unperforated sheet described above with that of a sheet with a
dense array of holes. We begin by removing the node $i=0$ situated in the
center of the sheet and all the nodes $j$ such that $r_{0j}=|\mathbf r_j| <
R$. We then repeat this operation periodically throughout the lattice to create
a dense lattice of perforations (see Fig.~\ref{fig:pattern} and Supplementary
Fig.~1). In this paper we consider arrays of holes of size $R=1,2$ with
varying spacing. We kept the radius of the hole small relative to the
length of the sheet to minimize finite-size effects.

\begin{figure} \includegraphics[height=\linewidth,angle=270]{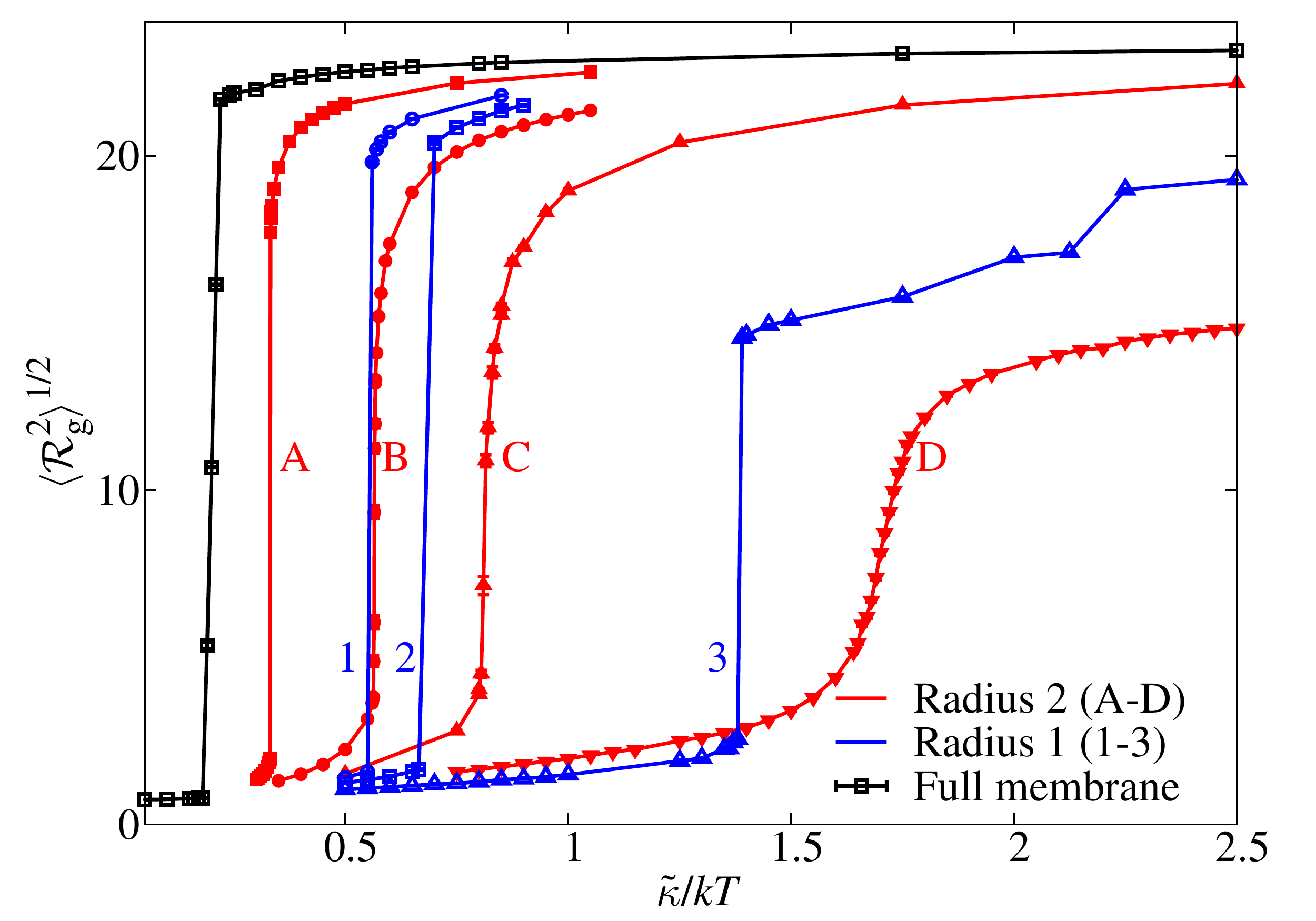}
\caption{Radius of gyration as a function of $\tilde\kappa/kT$ for several
patterns of perforation.  The black curve is the baseline unperforated
membrane, which crumples at the highest temperature,
$\tilde\kappa/kT_\text{c}\approx0.18$.  The red curves (A to D) are for
triangular arrays of perforations of radius $R=2$  with decreasing spacing
between individual holes.  The blue curves ($1$ to $3$) are arrays of
perforations with radius $R=1$.  As the spacing between holes is reduced, the
crumpling temperature decreases.  The full description of the perforation
patterns A--D and $1$--$3$ is given in Supplementary Fig.~1.  Data for
systems of size $L=100a$.  \label{fig:Rg-area} } \end{figure}

As a first demonstration of the dramatic effect of these perforations, consider
the pattern of holes depicted  in Fig.~\ref{fig:pattern}--right.
Fig.~\ref{fig:crumpling} compares the equilibrium configurations of this
perforated sheet and those of  the full membrane for two values of temperature
that differ by only $10\%$. In both cases, the full membrane is deep in the
flat phase and exhibits smooth, approximately flat configurations. The
perforated sheet, on the other hand, experiences a crumpling transition.

To characterize this transition it will be useful to consider the radius of
gyration of the sheet
 \begin{align}\label{eq:Rg} \mathcal R_\text{g}^2
&=\frac{1}{3N} \sum_{i=1}^N \langle\mathbf R_i\cdot \mathbf
R_i\rangle,& \mathbf R_i = \mathbf r_i - \mathbf r_\text{CM},
\end{align}
 where $\mathbf r_\text{CM}$ is the position of the center of
mass and $\langle O\rangle$ represents a thermal average.  In the flat phase,
$\mathcal R_\text{g}^2\sim L^{4/d_\text{H}}$, with Hausdorff dimension
$d_\text{H}=2$, while in the crumpled phase $\mathcal R_\text{g}^2 \sim
\log(L/a)$ ($d_\text{H}=\infty$). In the critical region, the Hausdorff
dimension has been computed with analytical methods ($d_\text{H} =
2.73$~\cite{LeDoussal1992}) and with numerical simulations
($d_\text{H}=2.70(2)$~\cite{Cuerno2016}).

\begin{figure}
\includegraphics[height=\linewidth,angle=270]{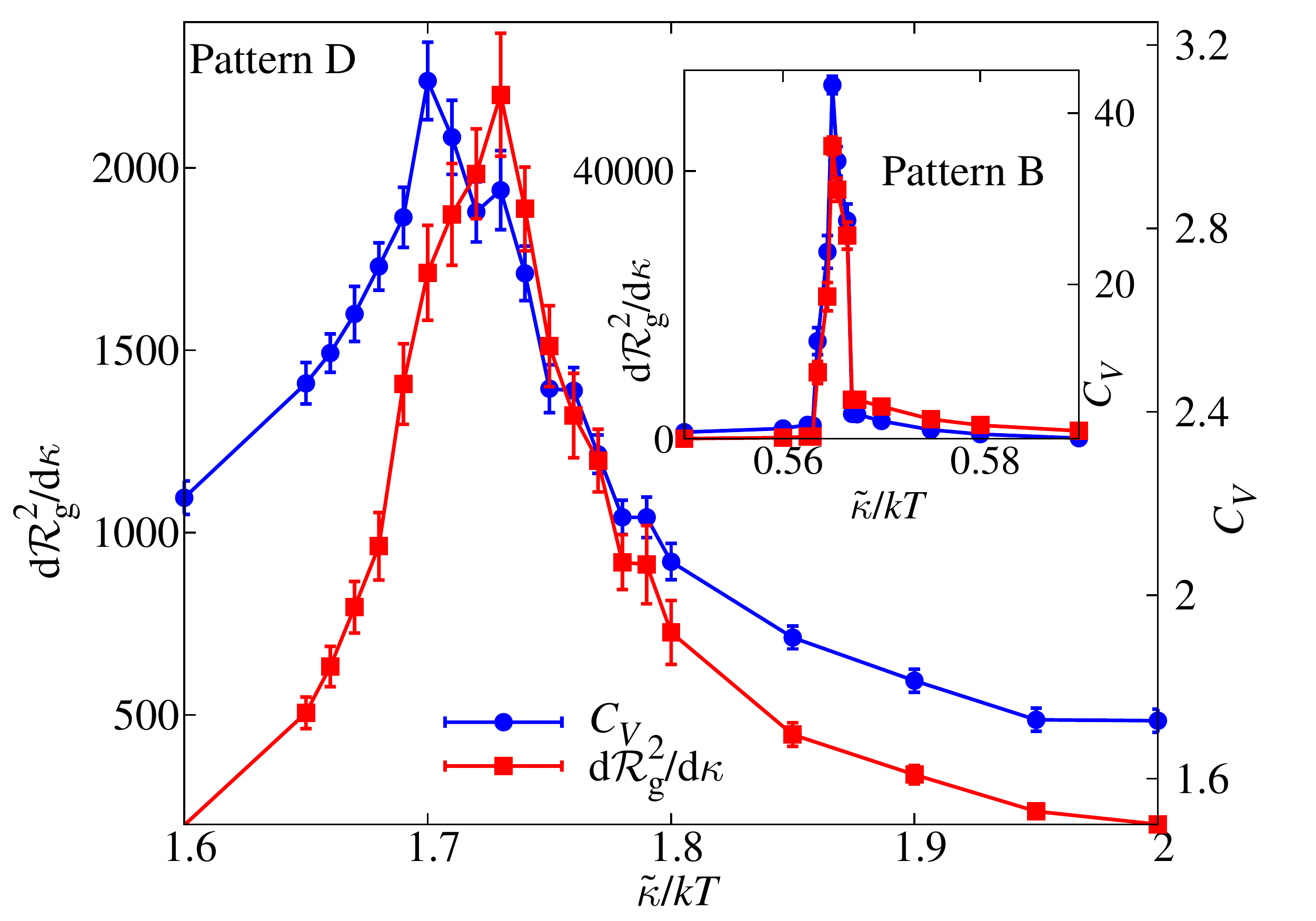}
\caption{Location of the crumpling temperature. We plot
the specific heat $C_V$ (right axis) and $\tilde\kappa$-derivative
of the radius of gyration (left axis) as a function of $\tilde \kappa/kT$ for
our most perforated system (corresponding to curve D in
Fig.~\ref{fig:Rg-area}).  The inset shows the analogous plot for a less
perforated sheet (corresponding to curve~B in Fig.~\ref{fig:Rg-area}), with a
much sharper transition (note the different vertical scales of the axes). All
error bars represent the standard error of the mean.
\label{fig:picos}
}
\end{figure}

We have plotted $\mathcal R_\text{g}$ as a function of $\tilde\kappa$ for all
our perforation patterns in Fig.~\ref{fig:Rg-area}.  In blue (red) we represent
systems with arrays of holes of radius $R=1$ ($R=2$) and a decreasing
separation between holes.  The black curve provides the baseline value of
$\mathcal R_\text{g}^2$ for the full membrane. We are interested in computing
the critical $kT_\text{c}/\tilde\kappa$ for crumpling in each of these
geometries. This can be done by searching for the maximum in the specific heat
of the system, which can be computed as~\cite{Harnish1991}
 \begin{equation} C_V
= \frac1N \bigl( \langle \mathcal H^2\rangle - \langle \mathcal
H\rangle^2\bigr)\ .  \end{equation}
 Alternatively, we can consider the
$\tilde\kappa$-derivative of $\mathcal R_\text{g}^2$, which can be evaluated
as:
 \begin{equation}\label{eq:DRg} \frac{\text{d} \mathcal
R^2_\text{g}}{\text{d}\tilde\kappa} = \frac{kT}{\tilde\kappa} \bigl( \langle
\mathcal H\rangle \langle \mathcal R_\text{g}^2\rangle
- \langle \mathcal H \mathcal R_\text{g}^2\rangle\bigr).  \end{equation}
 We show these two quantities for two different perforation patterns in 
Fig.~\ref{fig:picos}. In the thermodynamic limit, the position of the
peaks in $C_V$ and $\dd\mathcal R^2_\text{g}/\dd \tilde\kappa$   tend to the
same $kT_\text{c}/\tilde \kappa$ value. For our finite systems, we use the
difference in these peak positions for our most perforated membrane (the case
where the peaks are most separated) as an estimate of our systematic error in
$\tilde\kappa/kT_\text{c}$.

\begin{figure} \includegraphics[width=\linewidth]{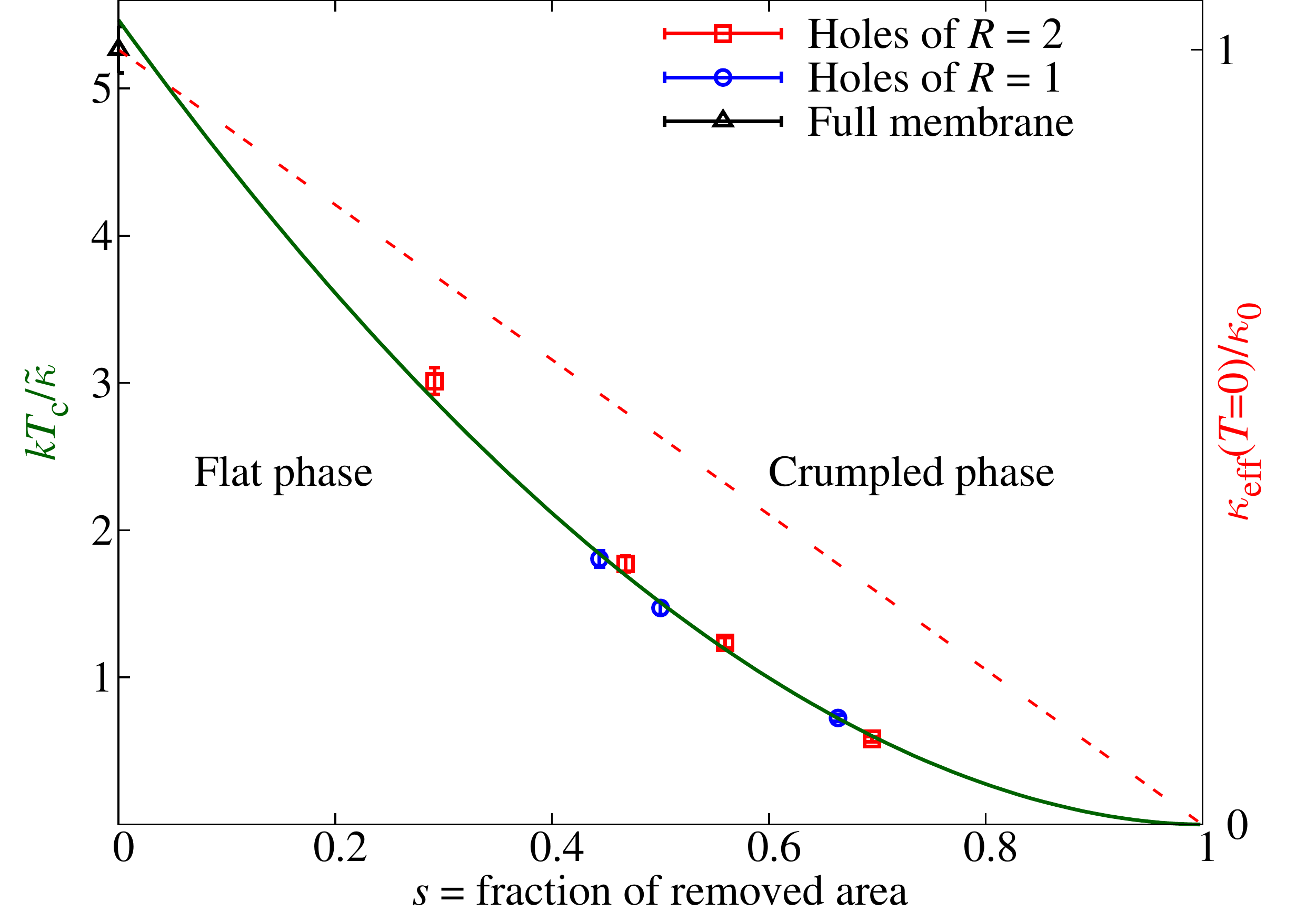}
\caption{Crumpling temperature $kT_\text{c}$ as a function
of the fraction $s$ of removed area. When plotted
against this parameter, the values of $kT_\text{c}$ for
all eight curves in
Fig.~\ref{fig:Rg-area} collapse to a single smooth function,
independent of the size of the
individual holes or their precise geometrical arrangement. The curve is a fit
to $f(s) = A(1-s)^{c}$, with $c=1.93(4)$ and a goodness-of-fit estimator
$\chi^2/\text{d.o.f.} = 8.72/6$ (d.o.f. = degrees of freedom). On the right-hand
vertical axis we also plot the zero-temperature effective bending rigidity
in units of $\kappa_0$ (red dotted line), which is simply linear in $(1-s)$.
The error bars represent an estimate of our systematic error, as explained
in the text.
\label{fig:kappa-dih}}  \end{figure}

In principle, one could think that this $T_\text{c}$ would depend in a
complicated way on the particular spatial arrangement of the holes or on their
individual sizes.  Fortunately, the reality is much simpler. Indeed, in
Fig.~\ref{fig:kappa-dih} we have plotted the $kT_\text{c}/\tilde\kappa$ for
each of the curves in Fig.~\ref{fig:Rg-area} as a function of the fraction of
removed area in the sheet. Given our discretization, this areal fraction is
most easily estimated by counting the fraction of remaining dihedrals
connecting adjacent triangles, after the holes have been made. As a function of
this dimensionless area fraction, all our $T_\text{c}$, including the one for
the full membrane, fall on a single smooth curve.
\begin{figure}
\includegraphics[height=\linewidth,angle=270]{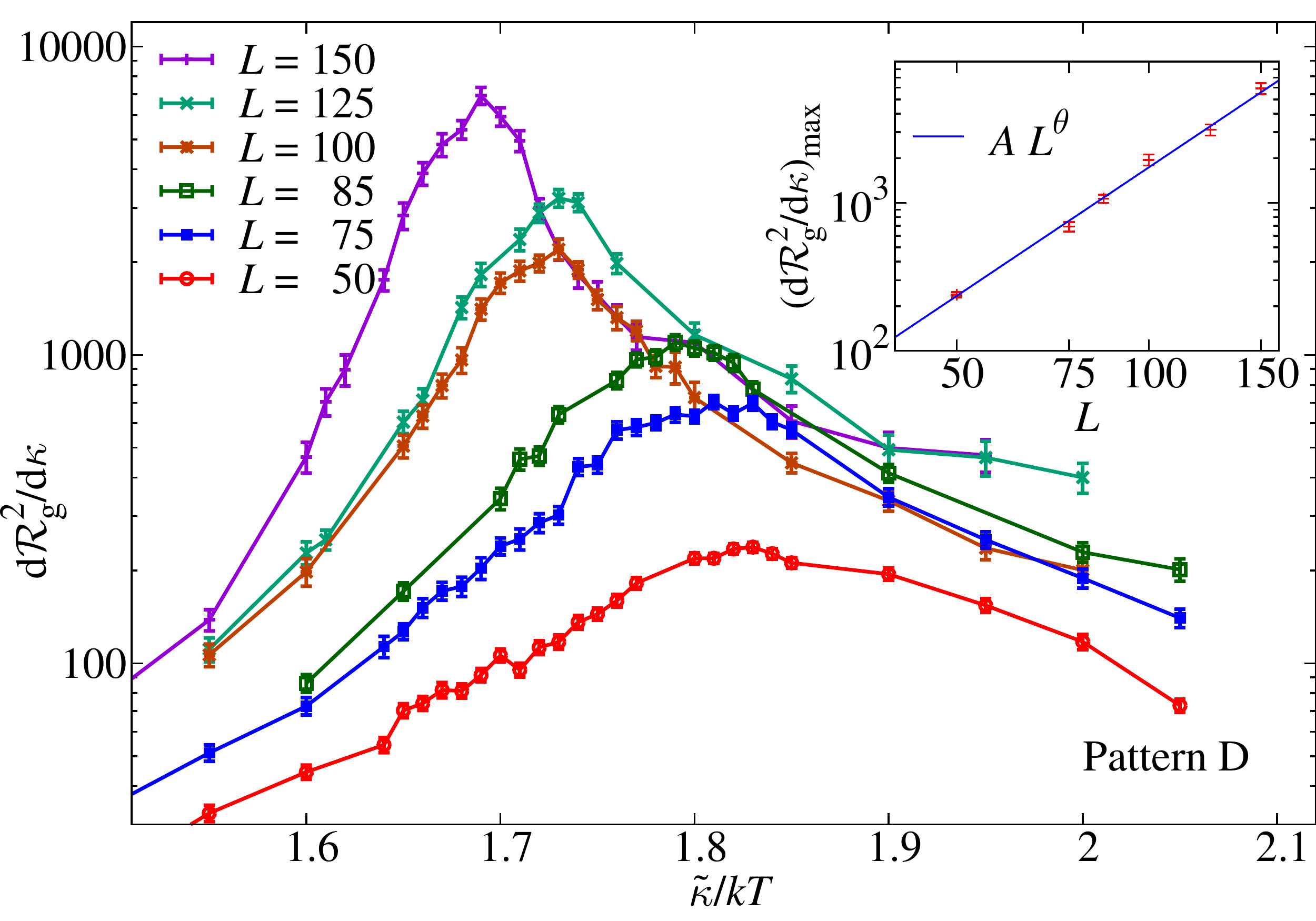} \caption{Peak of
$\dd\mathcal R_\text{g}^2/\dd\tilde \kappa$ for our most perforated sheet and
six system sizes $L$. Inset: scaling of the height of the peak with an exponent
$\theta=4/d_\text{H} + 1/\nu = 2.88(7)$, from a fit with $\chi^2/\text{d.o.f.} =
4.67/4$. The expected value for the crumpling transition~\cite{LeDoussal1992}
is $\theta\approx2.82$. All error bars represent the standard error of 
the mean. \label{fig:fss} } \end{figure}

In fact, if we denote by $s$ the fraction of removed area in the perforated
sheet, we have found that the following ansatz reproduces our results very
accurately: 
\begin{equation}\label{eq:kappac} k T_\text{c}/ \tilde \kappa = A
(1-s)^c.  \end{equation} 
With our choice of parameters, we obtain a good fit
with $c=1.93(4)$ and $A\approx5.5$. Notice, in particular, that for our most
perforated sheet (where about $70\%$ of the area has been removed), the value
of $kT_\text{c}$ is reduced by a factor of $10$ compared to the full membrane.
Extrapolating using Eq.~\eqref{eq:kappac}, we find that removing $85\%$ of the
area in a graphene sheet would bring the crumpling temperature down to about
$1600$~K.  Thus creating ``lacey graphene'' via, say, laser ablated holes that
remove $85\%$ of the carbon atoms could allow the crumpled regime to be
accessed experimentally. We note that the mechanical and electrical properties
of free-standing graphene springs with roughly $40\%$ of the material removed
were studied in~\cite{Blees2015}.

It is important to note that the observed $kT_\text{c}(s)$, \eqref{eq:kappac},
cannot be explained by the effective elastic constants of the perforated sheets
$\kappa_\text{eff}(T=0)$ and $Y_\text{eff}(T=0)$. Indeed, as we explain in
Supplementary Note~1 , the  $T=0$ bending modulus of the perforated sheet,
$\kappa_\text{eff}(T=0)$, linearly decays with $(1-s)$. Therefore, if the onset
of crumpling were simply determined by $\kappa_\text{eff}(T=0)$, one would
expect $T_\text{c}$ to be a linear function of $(1-s)$. Instead, as we obtained
in Eq.~\eqref{eq:kappac}, $T_\text{c} \sim (1-s)^{1.93}$, a result which
indicates that non-trivial thermal fluctuation effects are responsible.  The
effective Young's modulus $Y_\text{eff}(T=0)$, on the other hand, has a
complicated dependence on the details of the perforation pattern (see Supplementary
Note~1). However, $Y$
only affects the crumpling temperature as a logarithmic correction,
see Eq.~\eqref{eq:normals} below.  Explaining the observed value of $c=1.93(4)$
remains, therefore, a theoretical challenge.

\subsubsection*{Finite-size scaling} We have seen that cutting holes in a
membrane can induce crumpling at much lower temperatures. We have yet to show,
however, that this phenomenon quantitatively corresponds to the standard
crumpling transition that has been extensively studied for full
sheets~\cite{Kantor1987,Guitter1988,Renken1990,Harnish1991,LeDoussal1992,Bowick1996,
Wheater1996,Espriu1996,Koibuchi2004,Kownacki2009,Braghin2010,Hasselmann2011,Cuerno2016}.
This can be accomplished by  performing a finite-size scaling (FSS)
study~\cite{Amit2005} and finding the universality class of the phase
transition.  This computation poses two difficulties:
on the one hand our simulations cover a very wide range of temperature, rather
than concentrating all the numerical effort to increase the precision at the
critical region. On the other hand, the presence of the holes creates novel
finite-size effects. We begin by considering the FSS of the height of the peak
in $\dd\mathcal R_\text{g}^2/\dd\tilde \kappa$, which diverges
as~\cite{Cuerno2016} 
\begin{equation}\label{eq:dRgmax} \left. \frac{\dd
\mathcal R_\text{g}^2}{\dd \tilde \kappa}\right|_{\text{max}} \sim
L^{4/d_\text{H} + 1/\nu}\,, \end{equation} 
where $\nu$ is the critical exponent
describing a normal-normal correlation length that diverges at the crumpling
transition. When considering this equation, it is important
to notice that, while the exponent is universal, the algebraic prefactor is not and depends
on all the parameters.
In particular, for a given finite size, the transitions in Fig.~\ref{fig:Rg-area}
seem to be of varying sharpness. However, the values of the critical
exponents for the sharpest looking transition (the full membrane)
are known from previous work. In the following, we will perform the FSS analysis
and a fit to~\eqref{eq:dRgmax} only for our most perforated membrane (the 
rightmost curve in Fig.~\ref{fig:Rg-area}). If its critical exponents turn out
to be compatible with those of the full sheet, we can conclude that the 
intermediate curves will be in the same universality class too.
\begin{figure*}
\includegraphics[width=1.05\linewidth]{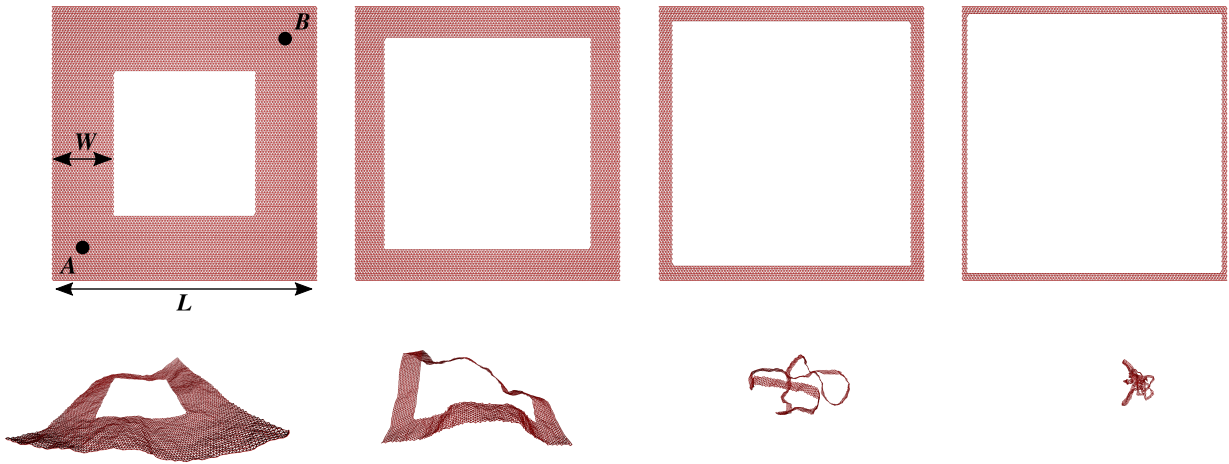}
\caption{Crumpling of a thin frame. The top row shows the initial ($T=0$) configuration
for frames of $L=100a$ and $W=24a,12a,6a,3a$ (left to right).
The bottom row shows thermalized configurations (for $\tilde\kappa=1.25kT$ and $\epsilon= 1800 kT/a^2$)
for each of these geometries, showing a clear crumpling as the frame width $W$ is reduced.
Points $A$ and $B$ of the leftmost frame are used to define an order parameter for crumpling in the text.
\label{fig:frames}}
\end{figure*}

Fig.~\ref{fig:fss} shows the results of this analysis.  We obtain $\theta=4/d_\text{H} + 1/\nu = 2.88(7)$, to be
compared with $\theta=2.86(1)$ from a recent dedicated FSS study for the full
membrane~\cite{Cuerno2016}.  Extracting the values of $d_\text{H}$ and $\nu$
separately is more difficult.  In principle, one could compute $\nu$ by
studying the drift in the position of the peak $T_\text{c}^{(L)}\simeq
T_\text{c}^\infty + A L^{-1/\nu}$, but this has very strong corrections to the
leading scaling~\cite{Cuerno2016}. Alternatively, one could consider the
critical scaling of the specific heat (yielding $\alpha/\nu$ and hence $\nu$
from hyperscaling), but in this case one has to include an analytical
contribution that introduces an extra fitting parameter: $C_V = C_a +
AL^{\alpha/\nu}$.  Since, unlike for the full membrane~\cite{Cuerno2016}, we
have to discard sizes $L<50$ due to finite-size effects, we do not have enough
degrees of freedom to obtain a reliable computation of $\nu$.  We have checked,
however, that the value $\nu=0.74$ for the standard crumpling transition is
consistent with our data (see Supplementary Note~2). Using this estimate
of $\nu$ we obtain $d_\text{H} = 2.62(7)$.  In short, the transition in these
perforated membranes is compatible with the universality class of the
crumpling transition for pristine sheets, even though its location is shifted
downward in temperature by an order of magnitude.
\subsubsection*{Crumpling of thin frames} 
\begin{figure}
\includegraphics[height=\linewidth,angle=270]{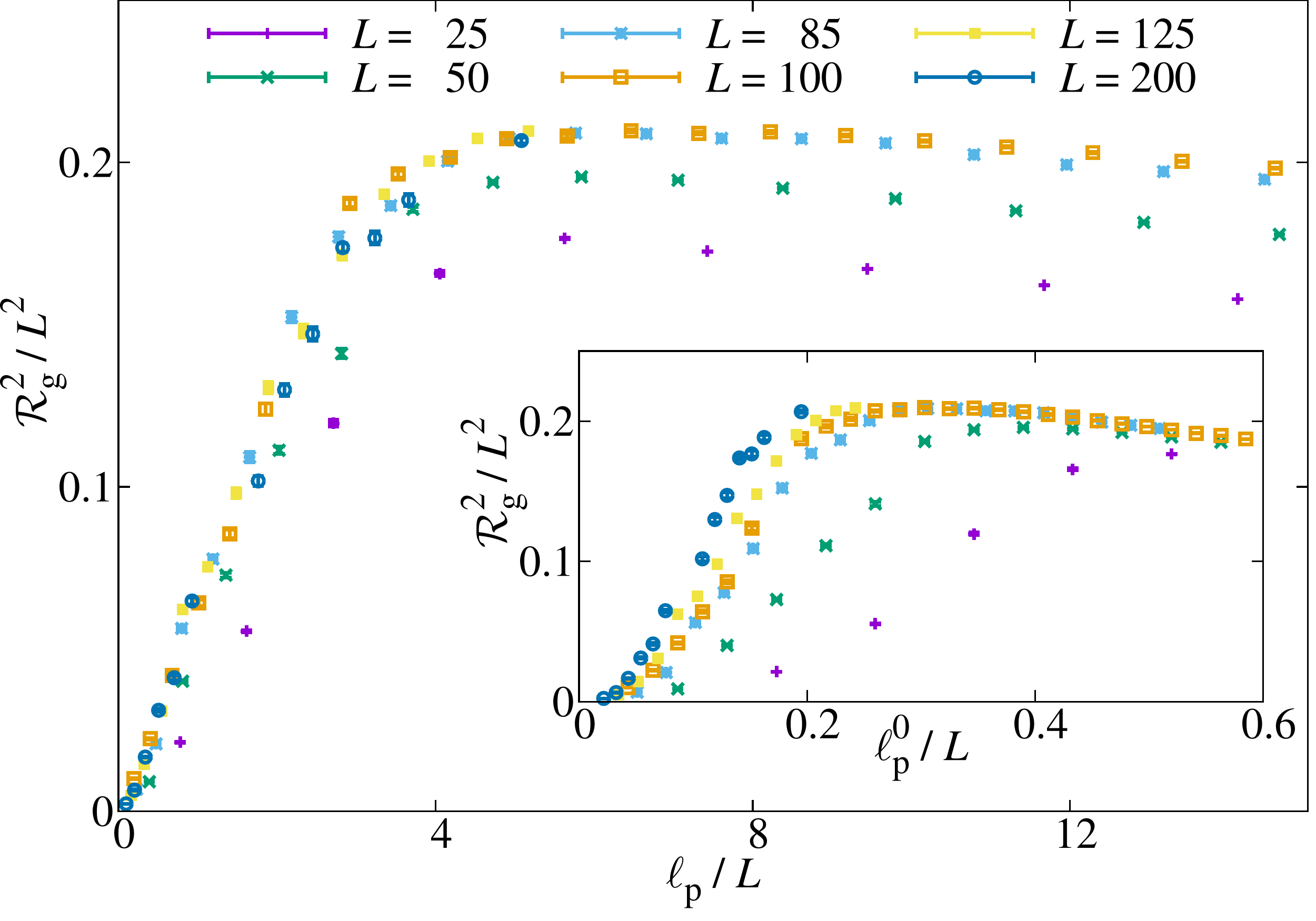}
\caption{Scaling in thin-frame crumpling. We 
plot the radius of gyration for frames of different $L$ and $W$
against $\ell_\text{p}/L$, where the persistence length $\ell_\text{p}=2W\kappa^\text{R}(W)/kT$ 
and the renormalized bending rigidity $\kappa^\text{R}(W)$ 
are defined in~\eqref{eq:Lp}. The curves for different system sizes collapse
when plotted against this scaling variable. The inset shows 
that neglecting thermal renormalization of the bending rigidity,
that is, considering $\ell_\text{p}^0 = 2W\kappa_0/kT$, leads to a poorer collapse.
In all these simulations we have used $\tilde\kappa = 1.25kT$ and $\epsilon=1800kT/a^2$.
All error bars (not visible at this scale in most cases) represent the standard
error of the mean.
\label{fig:Lp}}
\end{figure}

It is illuminating to consider what
happens when all perforations are combined to create a thin frame of width $W$
and overall size $L$, e.g., a membrane interrupted by a single large square
hole.  As shown in Fig.~\ref{fig:frames} (simulations at fixed temperature and
$L$ with varying $W$), there is now a striking crumpling transition as a
function of hole size. As an order parameter for this crumpling transition,
imagine erecting the normal to these frames at the points $A=(W/2,W/2)$ and $B=(L-W/2,L-W/2)$, where we
use an $xy$-coordinate system superimposed on the frame at $T = 0$ with origin
at the lower left corner. Then, in the flat phase of the frame (left side of
Fig.~\ref{fig:frames}, when the hole is small), we expect in the limit of large
frame sizes, $\langle\hat{\mathbf n}_A\cdot \hat {\mathbf n}_B\rangle \neq 0$.
Indeed, in the limit of a vanishingly small hole ($W\to L/2$), we expect~\cite{Kosmrlj2016}
\begin{equation}\label{eq:normals}
\langle\hat{\mathbf n}_A\cdot \hat {\mathbf n}_B\rangle_L = 1 - \frac{kT}{2\pi\kappa_0}
\left[ \eta^{-1} + \log\left(\frac{\ell_\text{th}}{a}\right) + C \frac{kT}{\kappa_0} \left(\frac{\ell_\text{th}}{L}\right)^\eta\right]\, ,
\end{equation}
where $C$ is a positive constant of order unity, $\eta\approx0.8$ and the thermal length scale is
\begin{equation}
\ell_{th} = \sqrt{\frac{16\pi^3\kappa_0^2}{3 kT Y_0}}\, .
\end{equation}
Thus $\lim_{L\to\infty}\langle\hat{\mathbf n}_A\cdot \hat {\mathbf n}_B\rangle_L\neq 0$, indicating that the normals
on diagonally opposite corners are correlated.
In contrast, when the frame is crumpled (right side
of Fig.~\ref{fig:frames}, when the hole is large), we 
clearly have $\lim_{L\to\infty}\langle\hat{\mathbf n}_A\cdot \hat {\mathbf n}_B\rangle = 0$.
 In the case of square frames, we can
estimate where the transition occurs by comparing the frame size $L$ to the
persistence length for thin frames of width $W$~\cite{Kosmrlj2016}.
\begin{align}\label{eq:Lp}
\ell_\text{p} &= \frac{2 W\kappa^\text{R}(W)}{kT}, & \kappa^\text{R}(W) &= \kappa_0 \left(\frac{W}{\ell_\text{th}}\right)^\eta.
\end{align}
 Here $\kappa^\text{R}(W)$ is the thermally renormalized bending rigidity.
Crumpling out of the flat phase should occur when $L>\ell_\text{p}$,
which suggests a scaling form for the radius of gyration of Eq.~\eqref{eq:Rg}, namely,
\begin{align}\label{eq:F}
\mathcal R_\text{g}^2 = L^2 F(\ell_\text{p}/L),
\end{align}
This scaling ansatz (where crumpling is indicated by the behavior for small $x$, $\lim_{x\to0} F(x) \sim x$)
is checked for a wide variety of frame dimensions $L$ as a function of $W$ in Fig.~\ref{fig:Lp},
which shows excellent data collapse as $L$ becomes large. Note that the
collapse is not nearly so good if one simply scales with a bare persistence
length (inset), indicating that thermal fluctuations play an important role in our
simulations. For this problem, it is known that the crumpled phase is robust to
distant self-avoidance~\cite{deGennes1979}. Indeed, the crumpled phase only
swells slightly with a scaling function in Eq.~\eqref{eq:F} that behaves accordingly
to $F(x) \sim x^{4/5}$ for small $\ell_\text{p}/L$. Of course, considerably more work would be required to demonstrate
convincingly that there is a sharp phase transition in the thermodynamic limit.
Here, the nontrivial width-dependent scaling of the thermally renormalized
persistence length in Eq.~\eqref{eq:Lp} suggests that the appropriate limit is $L,W\to\infty$ , with 
fixed $W (W/\ell_\text{th})^\eta/L$. In short, this analysis suggests that there could be a novel
transition for single frames, where both a crumpled and flat phase would survive in 
a polymer-like large-size limit. Even if this transition were simply a crossover, we expect a dramatic
change in mechanical properties, such as the response to bending, pulling and
twisting, when the frame crumples~\cite{Kosmrlj2016}.

We note finally that the crumpling temperature for unperforated membranes can be estimated (up to logarithmic 
corrections) from Eq.~\eqref{eq:normals} as $kT_\text{c} \approx 2\pi\eta \kappa_0$, in approximate agreement
with the transition temperature associated with the black curve in Fig.~\ref{fig:Rg-area}.

\begin{figure} \includegraphics[width=\linewidth]{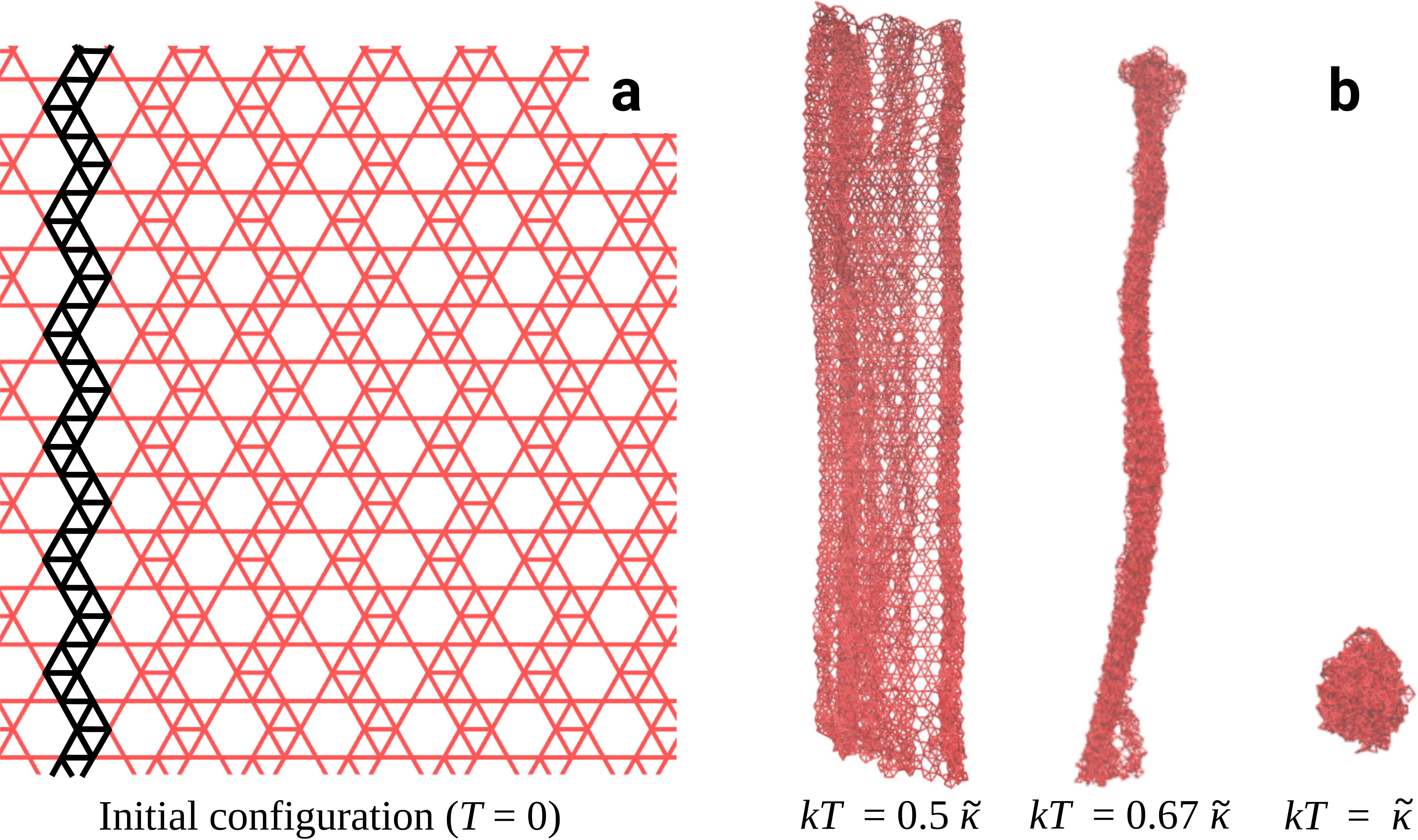} \caption{Two-step
crumpling transition in an anisotropic sheet. The zig-zag pattern of
approximately vertical struts reinforced by edge-sharing triangles make this
structure more rigid in the vertical than in the horizontal direction (see
highlighted example in the figure). We show snapshots of thermalized
configurations for several temperatures. As $T$ increases, the anisotropy in
the pattern of perforations makes the membrane first fold into a tight
cylinder, before crumpling completely.  This geometry corresponds to the system
labeled Pattern~3 in Fig.~\ref{fig:Rg-area} and in Supplementary Fig.~1.
The $T=0$ snapshot (\textbf{a}) is a close-up to a $30a\times30a$ section
of the lattice, while the finite-temperature snapshots (\textbf{b}) show the full
$100a\times100a$ system.
  \label{fig:aniso}} \end{figure}

\subsection*{Discussion} We have studied the mechanics of thermalized membranes
with a dense array of holes and found that the perforations can bring the
crumpling temperature into an experimentally accessible regime. From
Fig.~\ref{fig:kappa-dih}, we have $kT_\text{c}/\tilde\kappa\simeq A
(1-s)^{1.93}$ for the crumpling temperature as a function of the area fraction
removed $s$, independent of the detailed arrangement and size of our periodic
lattice of holes.  In addition, we have found that with an anisotropic pattern
of perforation one can induce a first partial crumpling at an even lower
temperature.  Indeed, see Fig.~\ref{fig:aniso}, a system where the perforations
are asymmetric or arranged in such a way that one of the membrane's axes
presents less bending resistance will first fold and roll into a very tight
cylinder, before crumpling completely. See reviews by Radzihovsky and by Bowick
in~\cite{Nelson2004} for a discussion of two-stage crumpling. These
observations provide a potential method for bridging the gap between the
theoretical expectations for the crumpling transition and the experimentally
accessible temperatures.

A  subtle issue is our neglect of distant self-avoidance.  The nearest-neighbor
springs in Eq.~\eqref{eq:H} embody an energy penalty of order $\epsilon a^2$
when nearest-neighbor nodes overlap, a number which greatly exceeds $kT$.
Adding a hard sphere excluded volume interaction between second-nearest
neighbors would create an entropic contribution to $\tilde\kappa$ of order
$kT$, which might produce a small shift in the crumpling temperature. The
existence of a sharp crumpling transition in unperforated membranes with
distant self-avoidance remains unclear at the present
time~\cite{Bowick2001,Wiese2001}. The presence of a lattice of large holes will
certainly reduce the effect of distant self-avoiding interactions, especially
when the removed area fraction becomes large. When distant excluded-volume
interactions are non-zero but weak, theory predicts a sharp transition between
a low-temperature flat phase and a high-temperature crumpled phase with a
nontrivial fractal dimension
$d_\text{H}\approx2.5$~\cite{Kantor1987c,Kardar1988,Paczuski1988}, qualitatively similar to
the findings for perforated membranes presented here. 
In addition, we have argued for the existence of a sharp crumpling transition when
all perforations are combined to create a thin frame with a single large hole
in the center of the membrane. In this case, it is well known that the crumpled
ring polymer phase survives the imposition of distant
self-avoidance~\cite{deGennes1979}. We hope our results will stimulate
allocation of resources (both experimental and computational) that will allow
investigations of distant self-avoidance in the presence of a lattice of
perforations. Even if distant self-avoidance smears out a sharp crumpling
transition, we nevertheless expect qualitatively different mechanical behavior
in the regimes identified here for thermalized kirigami sheets.

\subsection*{Methods}
\subsubsection*{Our simulations} We have simulated model~\eqref{eq:H}  for sizes ranging
from $L=25a$ to $L=150a$ with molecular dynamics in an $NVT$ ensemble, using a
standard Nos\'e-Hoover thermostat~\cite{Nose1984,Hoover1985}.  All  simulations
were carried out with the help of the HOOMD-blue
package~\cite{Glaser2015,Anderson2008}. Smaller sizes (up to $L=50a$) were
simulated on CPUs using  a message-passing interface (MPI) parallelization,
while for larger systems we have used GPUs. We use a simulation timestep of
$\Delta t=0.0025$ (in natural units where $a=m=kT=1$).  We start with a flat
sheet in the $xy$-plane,  and add a small random $z$ component to all the
nodes, in order to get the molecular dynamics started.  We then follow the
evolution for $2\times10^8$ timesteps, discarding the first $10\%$ for
thermalization and using a jackknife procedure~\cite{Amit2005} to estimate
statistical errors.  Converted into wall-clock time, $10^8$ steps of a
simulation of size $L=100a$ (with $11484$ nodes, $34023$ bonds and $33597$
dihedral angles) require about $8$~hours of execution time on an NVIDIA Tesla
K40. Our total simulation time has been the equivalent of $\approx5$ months of
a single Tesla K40. 

\paragraph*{Data availability.} The data that support the findings of this study are available from the corresponding
author upon reasonable request.

\subsection*{Acknowledgments} DY thanks Rastko Sknepnek for his guidance about
the HOOMD-blue package. This research was supported by the NSF through the
DMREF grant DMREF-1435794 and DGE-1068780, as well as  by the
Syracuse University Soft Matter Program. Work by DRN was also supported through NSF grants DMREF-1608501 
and via the Harvard Materials Science Research and Engineering Center, through
grant DMREF-1435999. We benefited from frequent discussions
with the experimental groups of P. McEuen and I. Cohen at Cornell University.
SB, MB and DY thank the KITP for hospitality during part  of this project and
Suraj Shankar for valuable discussions.  DY acknowledges funding by
through contract No. FIS2015-65078-C2-1-P, jointly funded by MINECO (Spain) and
FEDER (EU), and the
resources and assistance provided by BIFI-ZCAM (Universidad de Zaragoza), where
we carried out most of our simulations on the Cierzo supercomputer. 

\paragraph*{Author contributions.}  MB and DN designed the study; DY performed
the simulations and data analysis; SB and DY contributed analysis code; DY wrote
the paper with contributions from all authors.

\paragraph*{Competing interests.} The authors declare no competing financial
interests.

\appendix
\renewcommand{\figurename}{\textbf{Supplementary Figure}}
\setcounter{figure}{0}
\section*{Supplementary Figure 1}
\begin{figure}[h]
\includegraphics[width=\linewidth]{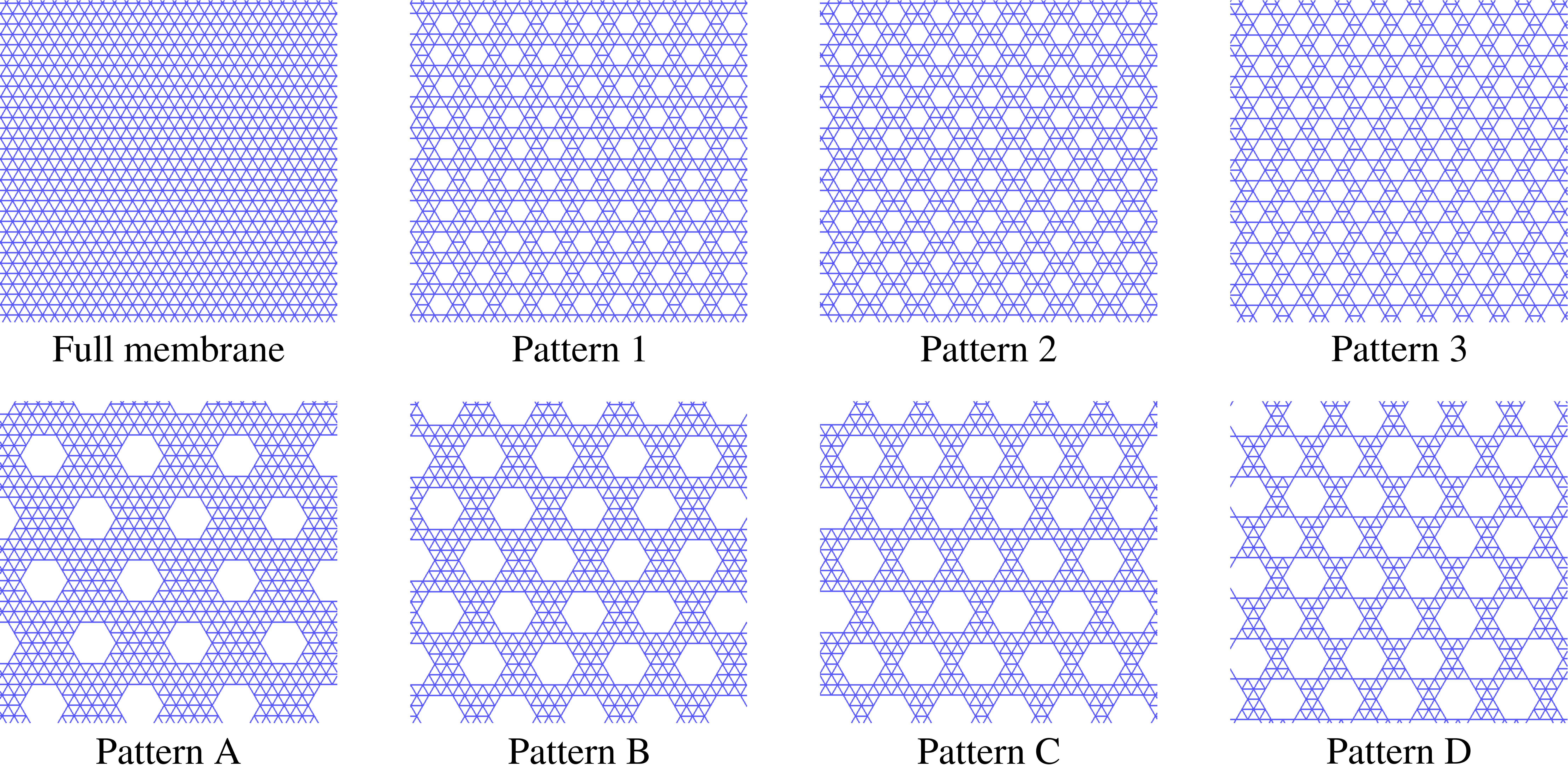}
\caption{
List of the membrane geometries considered in this
paper, using the same labels as in Fig.~3 in the main 
text. The top row shows the different perforation 
patterns with hole radius $R=1$ (plus the full membrane).
The bottom row shows the geometries wih
$R=2$. In order to see the details of these crystalline
membranes, we only show a $30a\times30a$ portion
of the full lattices (of size $L\times L = 100a\times100a$
throughout most of the paper).  We use our most perforated membrane (Pattern D)
for our finite-size scaling study.}
\end{figure}
\section*{Supplementary Note 1: \boldmath $T=0$ effective elastic constants of the perforated membranes}
\begin{figure}[h]
\includegraphics[width=\linewidth]{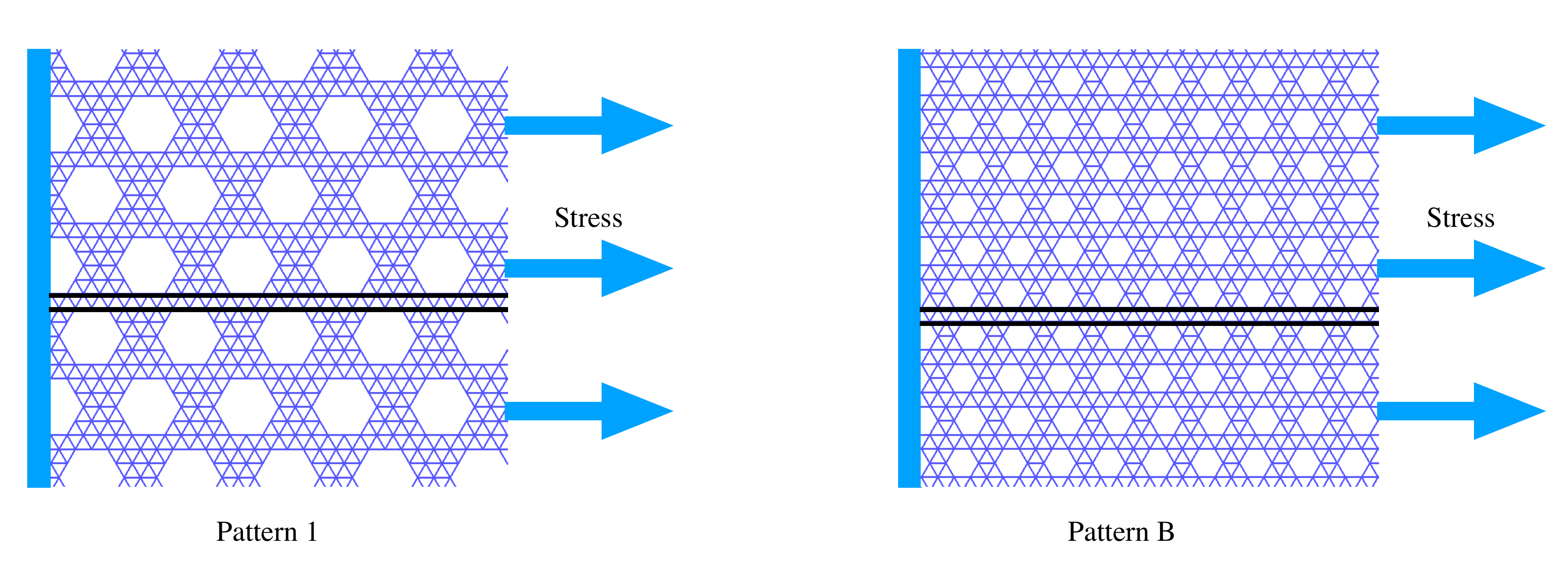}
\caption{Different responses to uniaxial stresses.
Diagram illustrating the different response of two patterns
with similar amount of removed area ($s=0.44$ for Pattern~1 and $s=0.47$ for Pattern~B).
For both cases we apply a uniaxial (horizontal) stress. Now, as one can see, 
in these $30a\times30a$ sections of the complete systems, Pattern~1 has~6 horizontal
linkages, while Pattern~B has~11 (we highlight one in each case). Therefore, we expect their Young's moduli
to be $11 Y_\text{eff}^{(1)} \approx 6 Y_\text{eff}^{(B)}$.  
 \label{fig:pattern}}
\end{figure}
The observed functional dependence of $T_\text{c}$ on the fraction of removed
area $kT_\text{c}/\tilde{\kappa} = A (1-s)^c$ cannot be explained just by evaluating
the $T=0$ elastic constants for perforated membranes. To see this, let us 
show that the $T=0$ bending rigidity is linear in the fraction of removed area.
We first note that the total energy at $T=0$ is a sum of positive-definite
terms. It is therefore always minimized by
setting all the springs to their rest length and the bending
energy to zero. The energy of a discretized
full sheet is $E_\text{bending}=E = \tilde \kappa N_\text{dih} E_0$, 
where $E_0$ is the energy of each dihedral and $N_\text{dih}$ 
is the  number of dihedrals. In order to write the energy in terms of the area,
we note that
the total area is just the sum of the areas of the individual triangles,
 $S=N_\text{tri} S_\text{tri}$. Now $N_\text{dih} = 3N_\text{tri}/2$,
so we can write $E= \tfrac32 \tilde \kappa N_\text{tri}E_0 =\tfrac32 \tilde \kappa E_0S/S_\text{tri}$. Taking 
the continuum limit we find $\kappa \propto E/S$. 
When we remove area the energy will decrease linearly in $S$,
which can be interpreted as a $\kappa_\text{eff}$ decreasing
linearly with $S$ while keeping the original intact area fixed.
If $\kappa_\text{eff}$ were to determine the crumpling temperature  
it would be linear in $(1-s)$ rather than decreasing as $(1-s)^{1.93}$.

We could instead consider the response of perforated membranes to
uniaxial stresses, which at $T=0$ will be characterized by a $Y_\text{eff}$.
In this case, we do expect the elastic response to be very sensitive 
to the details of the hole arrangement, leading to a non-linear dependence
of $Y_\text{eff}$ on the geometry.  We do not, however, expect $Y_\text{eff}$
to be relevant for the crumpling transition. Indeed, we know from the theory 
of thermalized elastic membranes that the Young's modulus
only affects $T_\text{c}$ as a logarithmic correction [see Eq.~(9) in the main text].

This can be illustrated with an example from our simulations
by looking closely at Patterns~1 and~B. Looking at Fig.~3., one will notice that
these two geometries have a very similar $\kappa/kT_\text{c}$.
Even though they look very different, 
counting the number of remaining triangles we see 
that the total fractions of removed area are very close ($s=0.44$ for Pattern~1
and $s=0.47$ for Pattern~B).  If we consider, on the other hand,
a stress applied in the horizontal direction at $T=0$, as in Supplementary Fig.~\ref{fig:pattern},
we see that their response will be significantly different.
Indeed, counting the number of horizontal linkages in each 
pattern we can estimate $11Y_\text{eff}^{(1)} \approx 6 Y_\text{eff}^{\text{(B)}}$,
a difference that does not reflect in their respective $T_\text{c}$. 

Notice, finally, that for anisotropic geometries such as Pattern~3 (Fig.~9),
the crumpling
transition is at the same temperature we would expect for an isotropic
pattern with the same removed area, further  reinforcing the result
that the elastic response to uniaxial stress is essentially irrelevant.
\newpage

\section*{Supplementary Note 2: The thermal critical exponent}
\begin{figure}[h]
\includegraphics[height=.6\linewidth,angle=270]{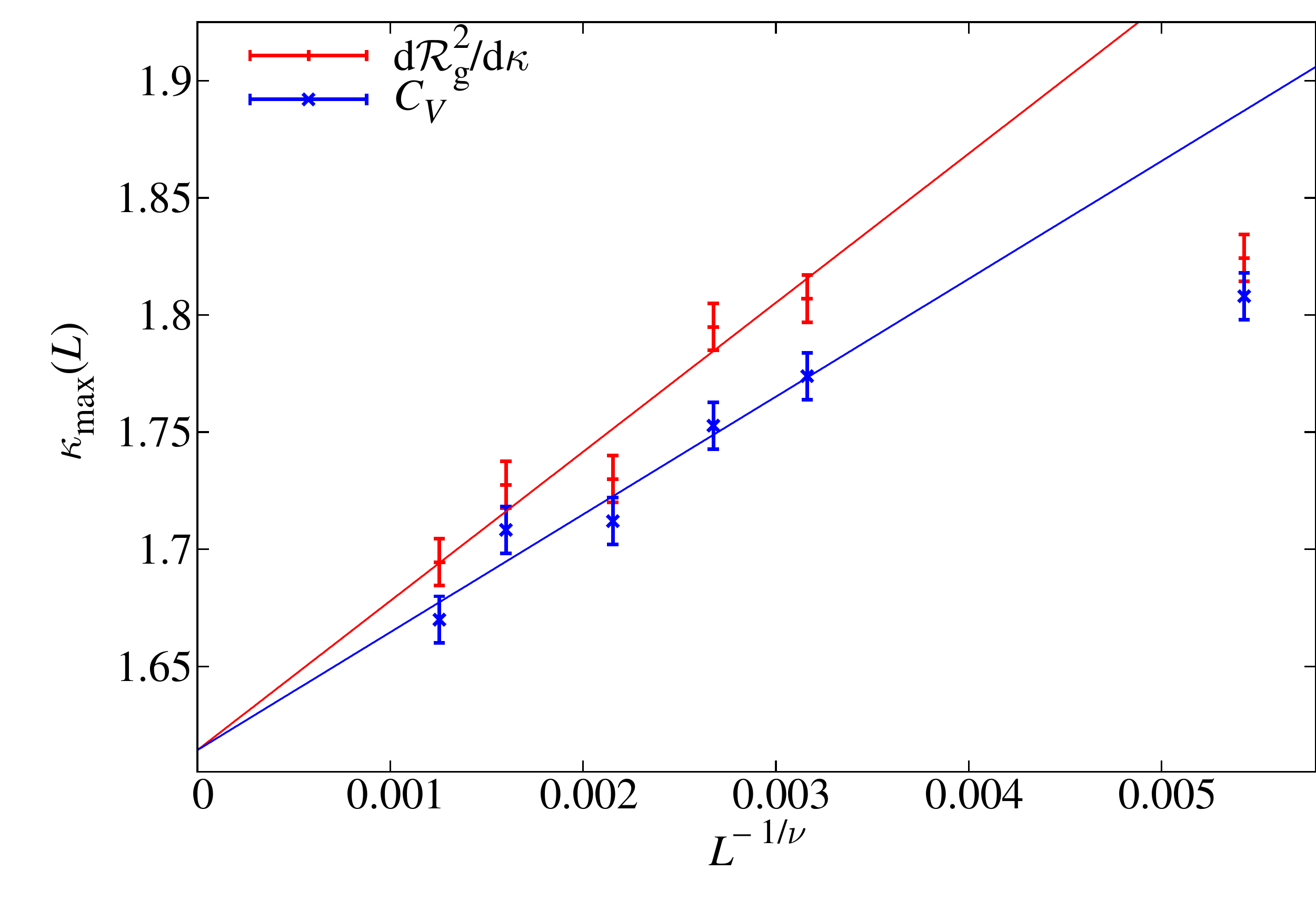}
\caption{The thermal critical exponent. We show the scaling of the position of the peaks
for the specific heat and the $\tilde\kappa$-derivative
of the gyration radius.  We plot the position of each
peak against $L^{-1/\nu}$, with $\nu=0.73$ from~\cite{LeDoussal1992}.
We then perform linear fits for $L\geq75$, forcing
both curves to extrapolate to the same $\tilde \kappa/kT_\text{c}$.
\label{fig:nu}}
\end{figure}
As discussed in the main text, a reliable 
direct computation of the thermal
critical exponent $\nu$ is not 
possible with our data. In principle,
one could study the evolution of the
position of the maximum for either
$C_V$ or $\dd\mathcal R_\text{g}^2/\dd\tilde \kappa$.
According to standard finite-size scaling~\cite{Amit2005}, this
quantity should shift with system size as 
\begin{equation}\label{eq:max}
\left(\frac{\tilde \kappa}{kT}\right)_\text{max} = \tilde\kappa/kT_\text{c} + A L^{-1/\nu}+ \ldots \, ,
\end{equation}
where the dots represent corrections to leading scaling for large $L$. Unfortunately, 
for this quantity these corrections are very strong (even for 
unperforated lattices~\cite{Cuerno2016}), so  in a fit
to~\eqref{eq:max} we would have to discard several system
sizes. This, together with the large fluctuations in the peaks' 
positions and the fact that we would have to fit 
both for the asymptotic value and for the exponent, 
prevents us from evaluating $\nu$ directly. We can, however,
verify that the known value for the standard crumpling transition ($\nu=0.73$~\cite{LeDoussal1992,Cuerno2016})
is compatible with our data. This is shown in Supplementary Fig.~\ref{fig:nu}.

\clearpage

\end{document}